\documentclass[twocolumn,aps,amsmath,amssymb,showpacs,superscriptaddress]{revtex4}

\usepackage{graphicx}
\usepackage{dcolumn}
\usepackage{bm}
\usepackage[latin1]{inputenc}
\usepackage[T1]{fontenc}
\usepackage[center]{subfigure} 
 \usepackage{tabularx}
\usepackage[hidelinks]{hyperref} 
 
\begin{document}

\title{\bf Apex predator and the cyclic competition in a  \\ rock-paper-scissors game of three species}
\author{C. A. Souza-Filho}
\affiliation{Departamento de F\'\i sica, Universidade Federal da Para\'\i ba, 58051-970, Jo\~ao Pessoa, PB, Brazil}
\affiliation{Instituto Federal de Educa\c c\~ao, Ci\^encia e Tecnologia da Para\'\i ba, Campus Princesa Isabel, 58755-000, Princesa Isabel, PB, Brazil}

\author{D. Bazeia}
\affiliation{Departamento de F\'\i sica, Universidade Federal da Para\'\i ba, 58051-970, Jo\~ao Pessoa, PB, Brazil}

\author{J. G. G. S. Ramos}
\affiliation{Departamento de F\'\i sica, Universidade Federal da Para\'\i ba, 58051-970, Jo\~ao Pessoa, PB, Brazil}

\pacs{87.10.Mn, 87.23.Cc}

\begin{abstract}
This work deals with the effects of an apex predator on the cyclic competition among three distinct species that follow the rules of the
rock-paper-scissors game. The investigation develops standard stochastic simulations but is motivated by a novel procedure which is explained in the work. We add the apex predator as the fourth species in the system that contains three species that evolve following the standard rules of migration, reproduction and predation, and study how the system evolves in this new environment, in comparison with the case in the absence of the apex predator. The results show that the apex predator engenders the tendency to spread uniformly in the lattice, contributing to destroy the spiral patterns, keeping biodiversity but diminishing the average size of the clusters of the species that compete cyclically.
\end{abstract}

\maketitle

\section{Introduction}

An intriguing problem in Biology concerns the understanding of how distinct species interact to maintain the mechanisms underlying biodiversity in nature. Several models focusing on the competing relations among species have been proposed and studied in the last few decades \cite{Frean_2001, Bacaer_2011, Adami_Nature2013}. In the simple case where the species compete for a single and restricted resource, their abilities appear to be hierarchical, involving a transitive relationship. In this case, one expects a winner species, leading all the other species to extinction \cite{Hardin_1960, Hutchinson1961}. However, there are other possibilities and when resource is abundant one may observe intransitive competing relationship, as it happens if one uses the rules of the rock-paper-scissors ($rps$) game, for instance. In this case, when one considers the system with three species, individuals of the species $A$ predate those of the species $B$, $B$ predates $C$ and $C$ predates $A$ in a cyclic competition environment. In this {\it rps} dynamics, all the species are treated equally and the system is known to lead to biodiversity \cite{Kerr_Nature2002, Karolyi_JTB2005, Reichenbach_Nature2007,Reichenbach_PRL2007, Nahum_2011,Sih_1985}. 

On the other hand, apex predators are being described as highly interactive from the biological point of view and their importance in the ecological environment has been the focus of several investigations \cite{Sih_1985, Palomares_1995, Terborgh_1999, Shurin_2001, Wallach_20152}.
The presence of an apex predator in a given ecosystem may favor coexistence of species, since it can diminish the process of competitive exclusion, imposing its own order to the set of species. This is known as predator-mediated coexistence \cite{Caswell_1978} and has been identified in several distinct environments, as in coral reef communities \cite{Paine_1966, Paine_1969, Porter_1972}, in communities of birds \cite{Kullberg_2000}, in vegetational diversity \cite{Jones_1933, Harper_1969, Paine_2002}, and in other scenarios.

Recent works also emphasize the use of the apex predator to restore ecosystems \cite{Ritchie_2012} that have been weakened due to distinct causes and motivations. In particular, one has identified secondary extinctions that could be maintained or restored in order to minimize damage or improve performance, as in the case of invasion of non native species \cite{Wallach_2010, Wallach_2015}, the transmission of diseases \cite{Pongsiri_2009} or the effects of climate changes on the dynamics of the species \cite{Wilmers_2006}. Such effects, which are triggered by the presence or action of the apex predator and may influence other levels of the chain, contributes to the so-called trophic cascade \cite{Silliman_2012}. Distinct factors such as changes in prey cascade behavior as a survival strategy and as predation that decreases the abundance of specific prey and interfere at distinc levels of the chain, may induce the occurrence of the trophic cascade.

In this work we investigate systems with three and four species and select three systems, one with three species that compete cyclically and two with four species, one with the four species competing cyclically and the other with the fourth species representing the apex predator, which predates all the other three species and is not predated by any of them. We focus mainly on the behavior of the apex predator and how it changes the behavior of the other species in the stochastic evolutions in the square lattice which we consider below. As we have commented above, the apex predador has been studied in several distinct scenarios, but the idea to be pursued in this work is original and of current interest. It develops standard stochastic simulations and was motivated by the recent work \cite{Bazeia_2017}, which makes use of the density of maxima related to the abundance of the species, and by the algorithm developed in Ref.~\cite{HK_1976}, which allows that we explore the clustering of species in the square lattice. 

The work is organized as follows. In the next section we introduce the four systems and study some specific time evolutions to show how they evolve under the rules there discussed. We move on and in Sec.~\ref{results} one selects and investigates three distinct systems, with the results of the stochastic simulations adding novelties to the current understanding of the behavior of the apex predator and how it changes the behavior of the other three species that compete cyclically. We end the work in Sec.~\ref{end}, where we include some comments and conclusions.

\section{The Models}
\label{models}

\begin{figure}[!ht]
\centering
\subfigure[]{\includegraphics[width=0.4\linewidth]{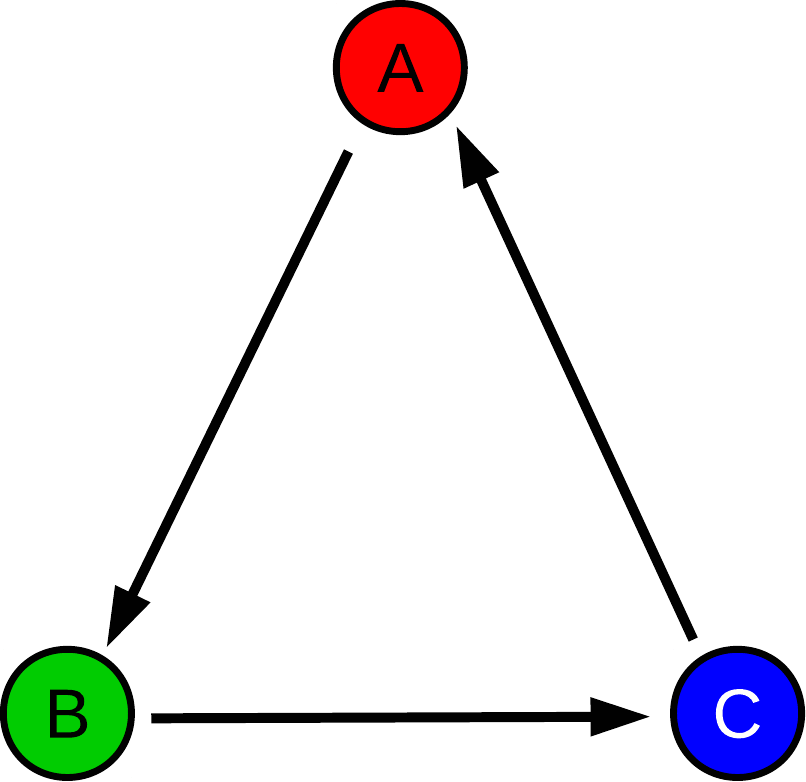} \label{foodweb}}\hspace{0.2cm} 
\subfigure{\includegraphics[width=0.4\linewidth]{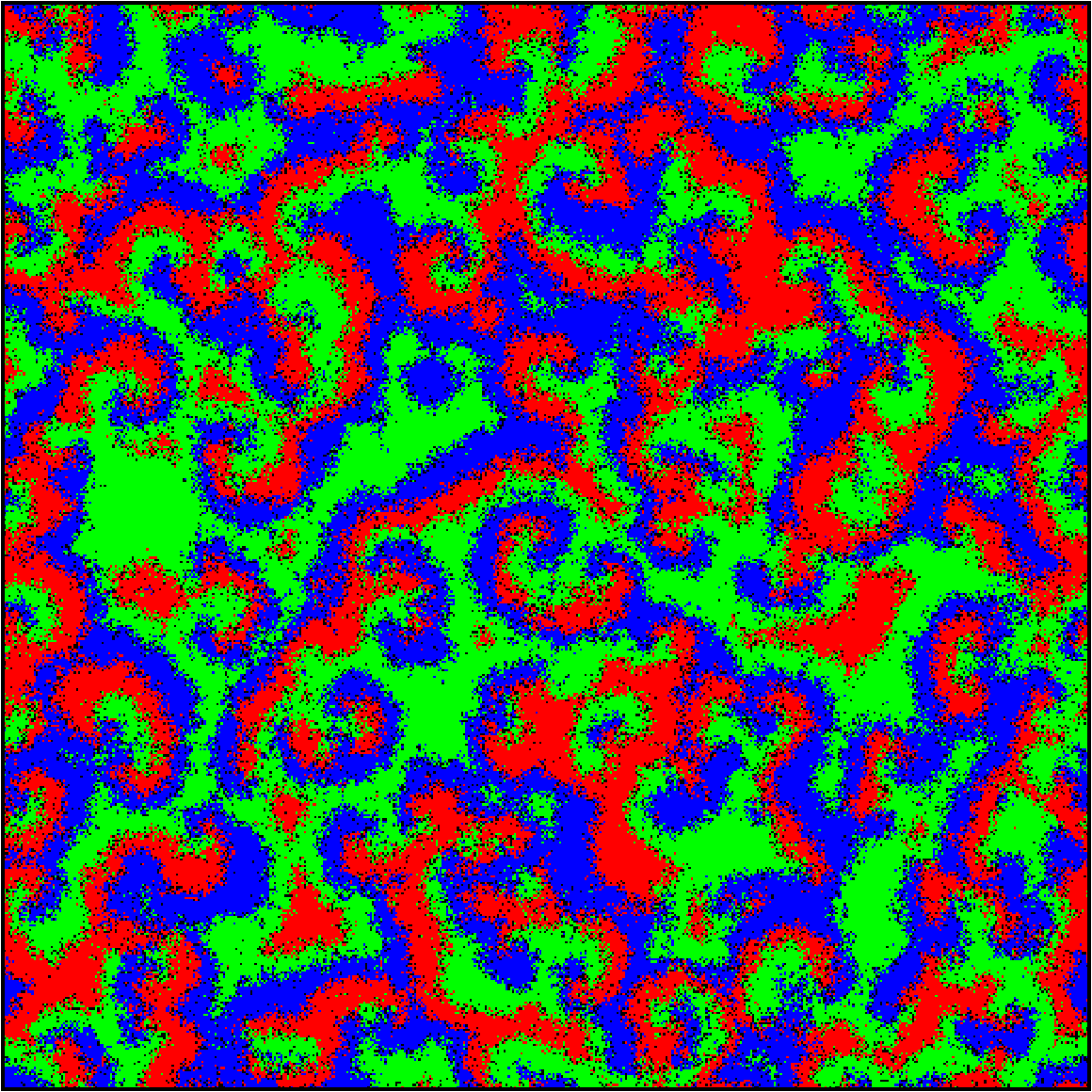} \label{rede_final}}\hspace{0.2cm}\\ \addtocounter{subfigure}{-1}
\subfigure[]{\includegraphics[width=0.4\linewidth]{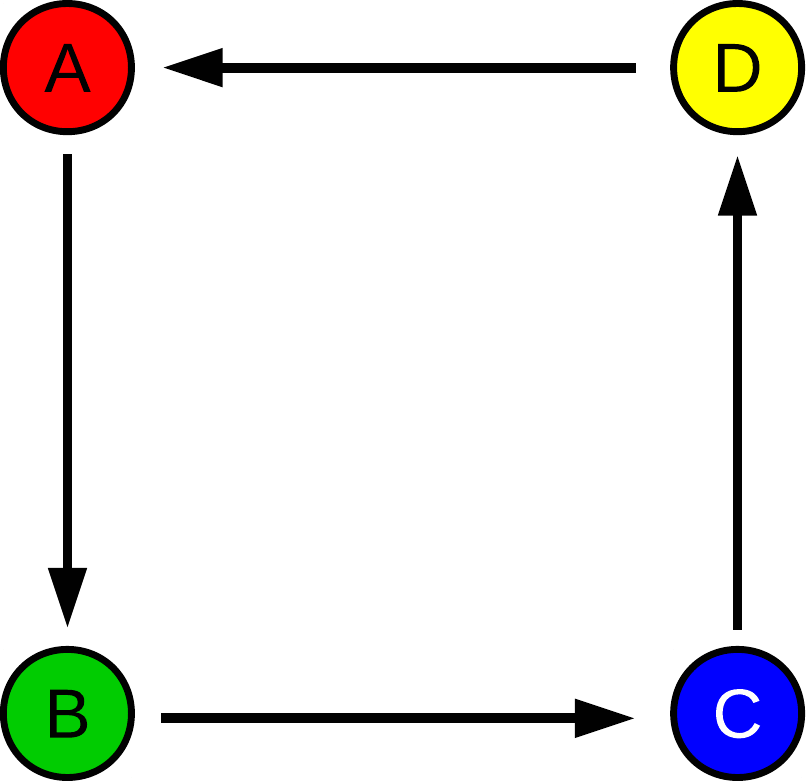} \label{foodweb1}}\hspace{0.2cm} 
\subfigure{\includegraphics[width=0.4\linewidth]{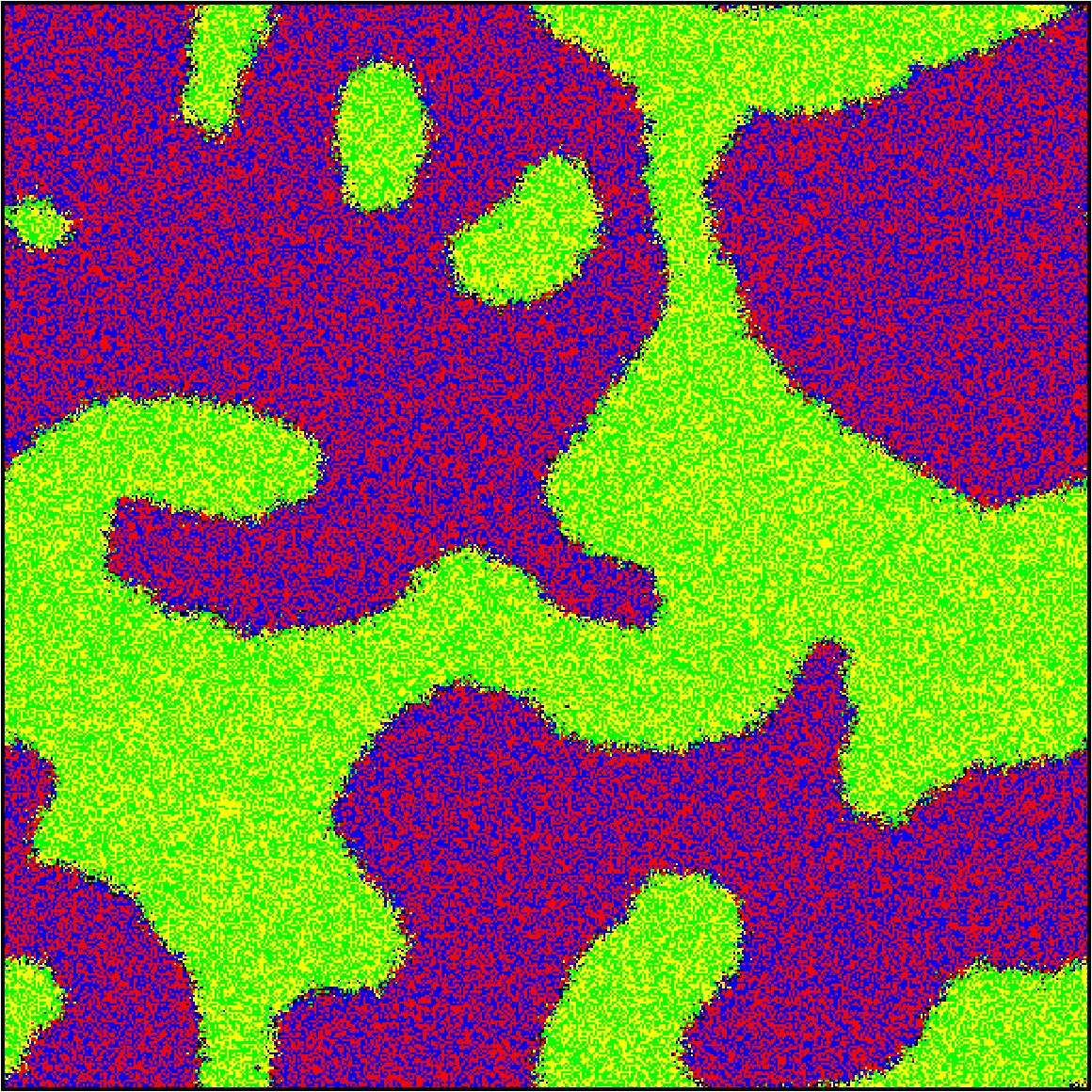} \label{rede_final1}}\hspace{0.2cm} \\ \addtocounter{subfigure}{-1}
\subfigure[]{\includegraphics[width=0.4\linewidth]{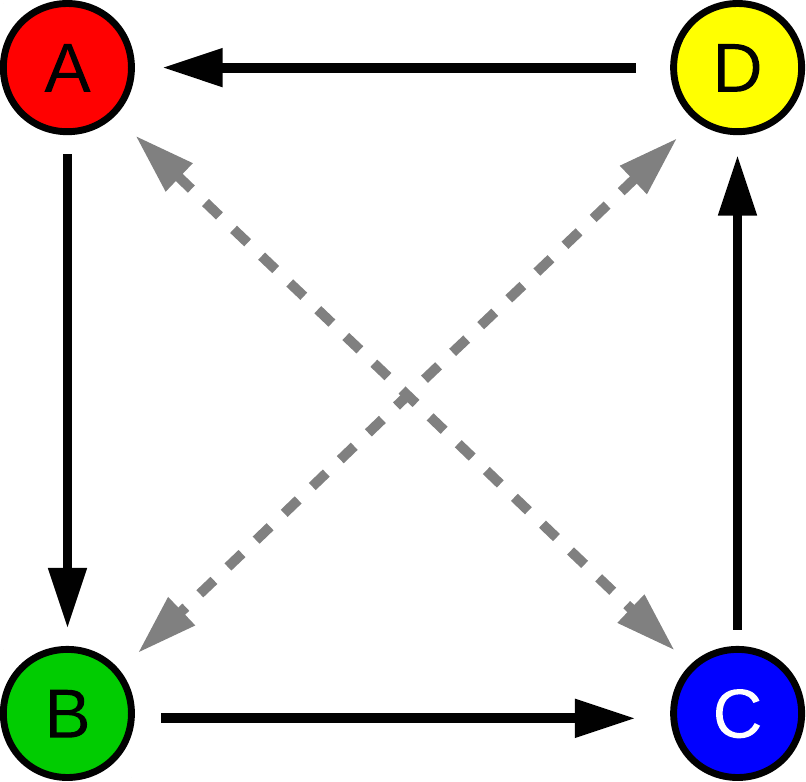} \label{foodweb2}}\hspace{0.2cm} 
\subfigure{\includegraphics[width=0.4\linewidth]{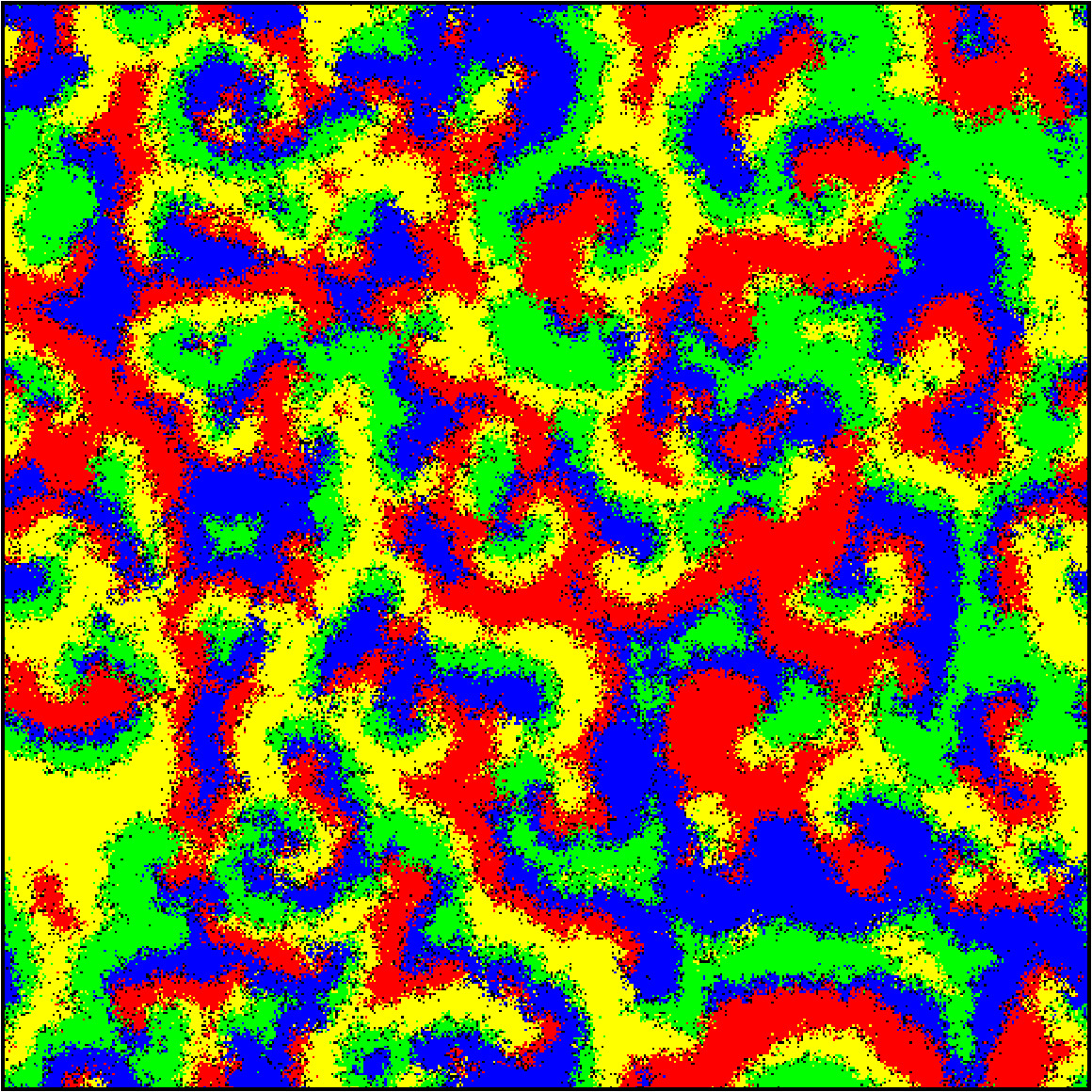} \label{rede_final2}}\hspace{0.2cm}\\ \addtocounter{subfigure}{-1}
\subfigure[]{\includegraphics[width=0.4\linewidth]{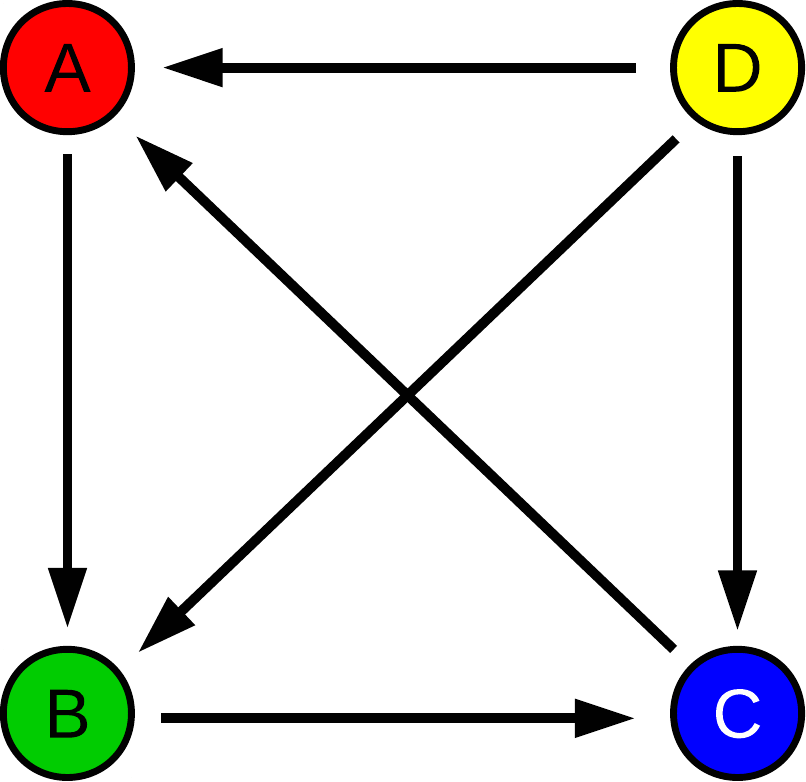} \label{foodweb3}} 
\subfigure{\includegraphics[width=0.4\linewidth]{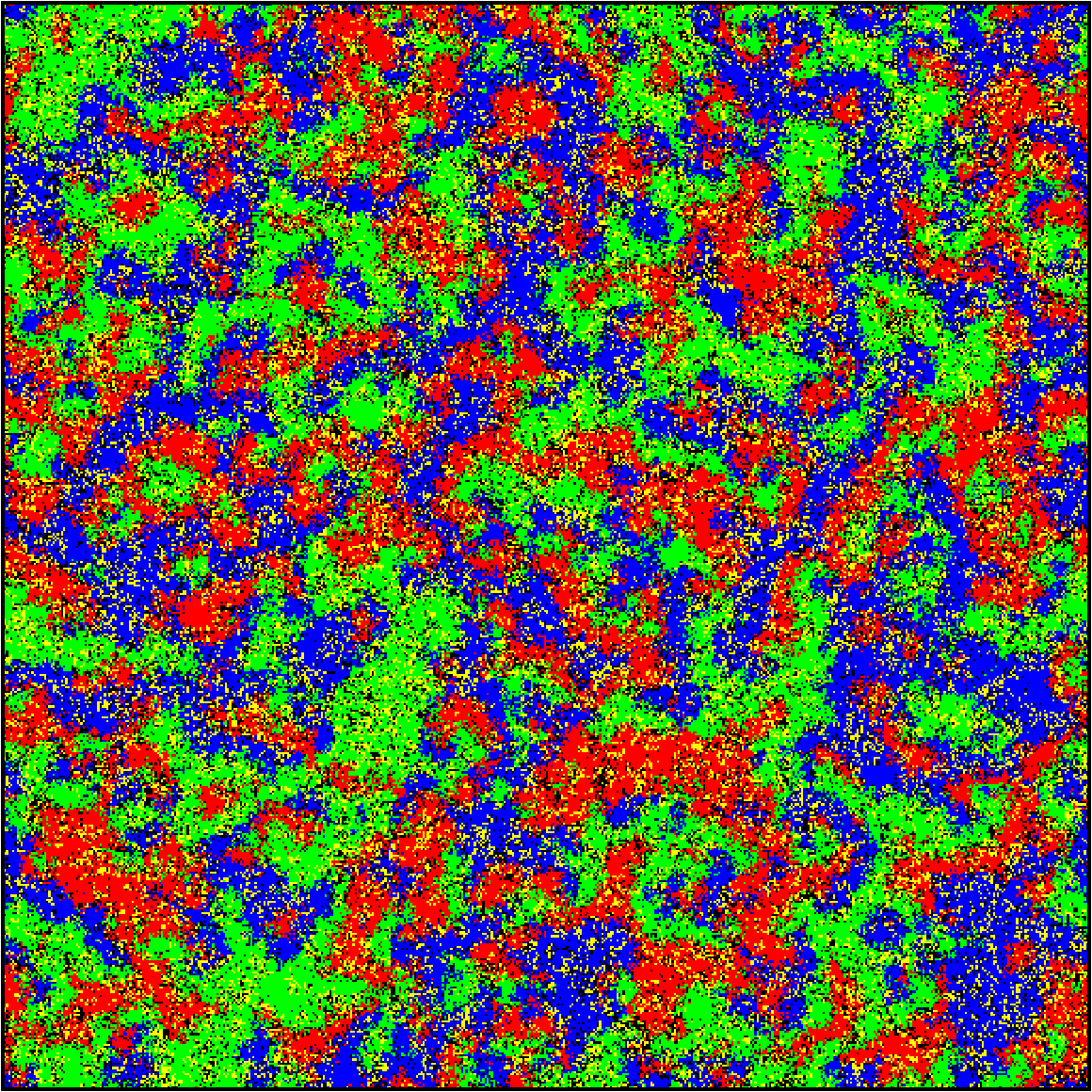} \label{rede_final3}}
\caption{In the left panel one displays the diagrams to show how predation works for the three and four species. The black arrows indicate unidirectional predation and the gray dashed arrows indicate bidirectional predation. In the right panel one displays typical snapshots of stochastic simulations that run for $10^4$ generations. In (d) one shows the system where the fourth species constitutes the apex predator.}
\label{foodwebs}
\end{figure}

In this work we investigate systems described by three and four species. In Fig.~\ref{foodwebs} one illustrates the systems and the interactions among the species, with the arrows going from the predator to the prey. In Fig.~\ref{foodwebs} (a) one shows three distinct species that compete cyclically, as in the rock-paper-scissors game. This is the system $X3$ and several studies were already implemented; see, e.g., Refs.~\cite{Peltomaki_2008, Wang_2010, Shi-Wang-Yang-Lai_PRE2010, Jiang-Zhou-Perc-Wang_PRE2011,Groselj_2015}. Among several interesting characteristics, one notes the presence of spiral patterns.
In Fig.~\ref{foodwebs} (b) one shows another system with four distinct species that also interact cyclically, but in this case the fourth species adds the next-to-next neighbor which does not compete and so form partnerships. This and other similar cases were studied in \cite{Szabo_2004, Szabo_2005,Intoy_2013, Lutz_2013, Avelino-Bazeia_PRE2012} and may generate patterns as the one shown in the figure, with two subsets of two partner species. In Fig.~\ref{foodwebs} (c) and (d) we show the systems $X4$ and $SX3$, respectively. They have all the four species interacting, but in (c) they are all equivalent, and in (d) one selects the yellow species to represent the apex predator.   

In order to implement the stochastic simulations, we consider a square lattice with $N=L^2$ sites and use periodic boundary conditions. We take $L=512$, so we deal with a square lattice of $512\times512$ sites. Each site in the lattice is occupied by one of the species $A$ (red), $B$ (green), $C$ (blue) or $D$ (yellow), or is empty $E$ (black). The interactions follows the rules
\cite{Reichenbach_Nature2007,Reichenbach_PRL2007}:

\begin{eqnarray}
AB \xrightarrow{\sigma_{\ }} AE,& \hspace{0.4cm} BC \xrightarrow{\sigma_{\ }} BE,& \hspace{0.4cm} CA \xrightarrow{\sigma_{\ }} CE,  \label{eq:cyclic}\\
AE \xrightarrow{\mu_{\ }} AA,& \hspace{0.4cm} BE \xrightarrow{\mu_{\ }} BB,& \hspace{0.4cm} CE \xrightarrow{\mu_{\ }} CC, \label{eq:reproduction}\\
A\square \xrightarrow{\varepsilon_{\ }} \square A,& \hspace{0.4cm} B\square \xrightarrow{\varepsilon_{\ }} \square B,& \hspace{0.4cm} C\square \xrightarrow{\varepsilon_{\ }} \square C,\label{eq:migration}
\end{eqnarray}
where $\square$ represents a site that can be empty or occupied by any individual. The relations in (\ref{eq:cyclic}) describe predation, which is characterized by the $\sigma$ parameter that shows the cyclic interactions. The relations in (\ref{eq:reproduction}) and (\ref{eq:migration}) show reproduction and migration, which occur controlled by the $\mu$ and $\varepsilon$ parameters, respectively. 

In the last system in Fig.~\ref{foodwebs} (d), the D or yellow species interacts obeying the following rules
\begin{eqnarray}
DX &\xrightarrow{\gamma_{\ }}& DD, \label{eq:agressividade}\\
D\,\square &\xrightarrow{\beta_{\ }} &E\,\square, \label{eq:morte}\\
D\,\square &\xrightarrow{\epsilon_{\ }}& \square\, D,  \label{eq:troca}
\end{eqnarray}
where $X$ represents one among the three species $A$, $B$, or $C$ that compete cyclically. These rules show that the D or yellow species describes the apex predator. The relation (\ref{eq:agressividade}) ensures that it can reproduces after predating any of the three species under the same ratio $\gamma$, and the relations (\ref{eq:morte}) and (\ref{eq:troca}) describe dead and migration, which are controlled by the parameter $\beta$ and $\epsilon$, respectively. 

Before starting the simulation one prepares the initial state, in which all the species and empty sites are evenly distributed in the square lattice with the same probability. Each time step randomly selects a site and one of its four nearest neighbors, and for each selected pair, the random process continues using the normalized ratios controlled by $\sigma$, $\mu$, $\varepsilon$, $\gamma$, $\beta$, and $\epsilon$, as described by the above processes \eqref{eq:cyclic}-\eqref{eq:troca}. To describe the time evolution of the system, we use generation, which is the time spent to account for $N$ time steps. 

In the right panels in Fig.~\ref{foodwebs} (a) and (c), one sees the appearance of spiral patterns, which is typical of the cyclic evolutions that follow the rules of the rock-paper-scissors game. However, in Fig.~\ref{foodwebs} (d) one notices the absence of spirals and the diminishing of the size of the clusters of the species that evolve in cyclic competition. This behavior is due to the presence of the apex predator, and we study it below, showing that although the apex predator does not destroy biodiversity, it diminishes the average size of the clusters of species that compete cyclically. We have added in \url{https://goo.gl/QexPxf} three videos to illustrate how the three systems $X3$, $X4$, and $SX3$ evolve in time under the stochastic simulations. 

\section{Results}
\label{results}

Let us now implement the stochastic simulations, taking the parameters for predation, reproduction and migration in Fig.~\ref{foodwebs} (a) and (c) as $\sigma = 0.25$,  $\mu = 0.25$ and $\varepsilon = 0.50$, respectively. For the system with the apex predator which appears in Fig.~\ref{foodwebs} (d), one uses $\sigma = 0.30$, $\mu = 0.30$, $\varepsilon = 0.40$, $\gamma = 0.25$, $\beta = 0.15$ and $\epsilon = 0.60$.
We have checked that the results in this work are robust against changes in the values of the parameters, if they are chosen in a way that maintain coexistence among the species. The main focus of the current work is to investigate the systems shown in Fig.~\ref{foodwebs} (a), (c), and (d), which we refer to as $X3$, $X4$, and $SX3$, respectively. 

\begin{figure}[!ht]
\centering
\includegraphics[width=1.0\linewidth]{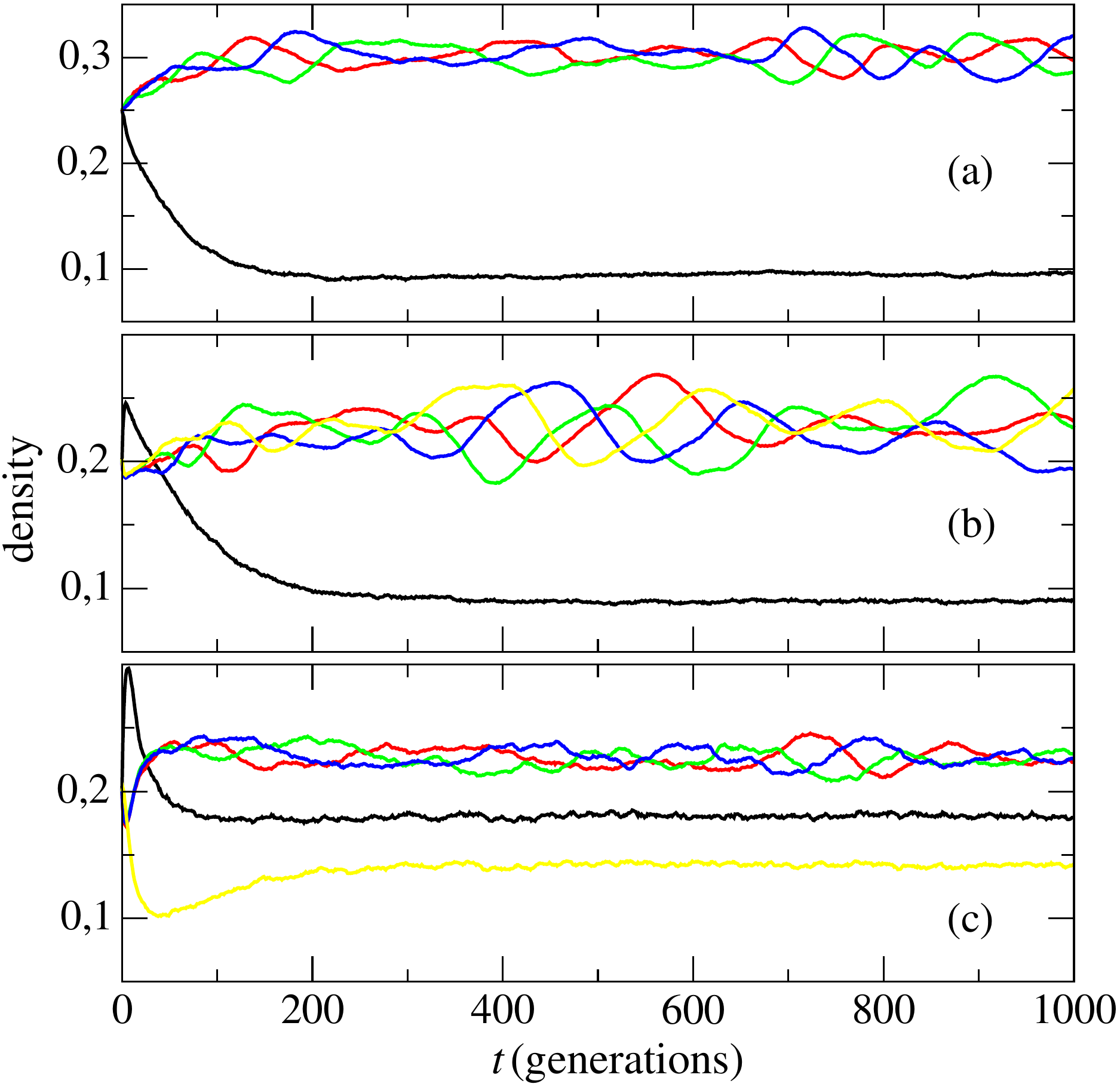} 
\caption{Density of species as a function of time. The colored curves represent the respective species, and the black curve stands for the empty sites. The (a) and (b) panels represent the $X3$ and $X4$ systems, and the (c) panel stands for the last system, $SX3$, which contains the apex predator. The long time evolution shows that all the systems evolve ensuring coexistence of species.}
\label{densidades}
\end{figure}

In Fig.~\ref{densidades} one displays the density of species, $\rho_x(t)$, with $x \in \{A,B,C,D\}$, as a function of the generation time. The colored curves represent the corresponding species, and the black curve identify the empty sites. The cases displayed in Fig.~\ref{densidades} (a), (b) and (c) represent the systems $X3$, $X4$ and $SX3$ respectively, and one notices that after a transient time interval, all the systems evolve maintaining coexistence of the species.

One sees that in the absence of the apex predator, the density of empty sites diminishes rapidly. This happens because the initial density of species and empty sites favor the reproduction, since the species can frequently meet empty sites. In the presence of the apex predator, the density of empty sites increases before diminishing rapidly, and now the reason is that the presence of empty sites contributes to the death of the apex predators, before they starts to increase to reach an equilibrium state that oscillates around a given average. 

In all the above cases, the density of species oscillates around an average value. This behavior is expected, since the systems are similar to the famous predator-prey model of Lotka-Volterra \cite{L,V}; see, e.g., Fig.~6 of Ref.~\cite{bazeia}, which shows similar mean field and stochastic network simulations in a 5 species model in the absence of the apex predator.
However, in the presence of the apex predator which we investigate in the current work, the frequency of oscillation of the density of species increases, with the decreasing of the amplitude. This occurs because the apex predator acts uniformly in the square lattice, and is expected to equally suppress all the species.
The average value of the density of species depends on the values used for the parameters that we consider in each case; for instance, the value obtained for the $X3$ system was $\overline{\rho}_x = 0.301$, with
$\sigma_x = 0.012$, where $\sigma_x$ is the standard deviation, and for the system $X4$, $\overline{\rho}_x = 0.225$, with $\sigma_x = 0.018$.
For the system $SX3$, we obtained $\overline{\rho}_x = 0.226$ with $\sigma_x = 0.006$ for the three species ${A,B,C}$ that evolve in cyclic competition, and $\overline{\rho}_d = 0.142$ with $\sigma_d = 0.001$, for the apex predator.

\begin{figure}[!ht]
\centering
\includegraphics[width=\linewidth]{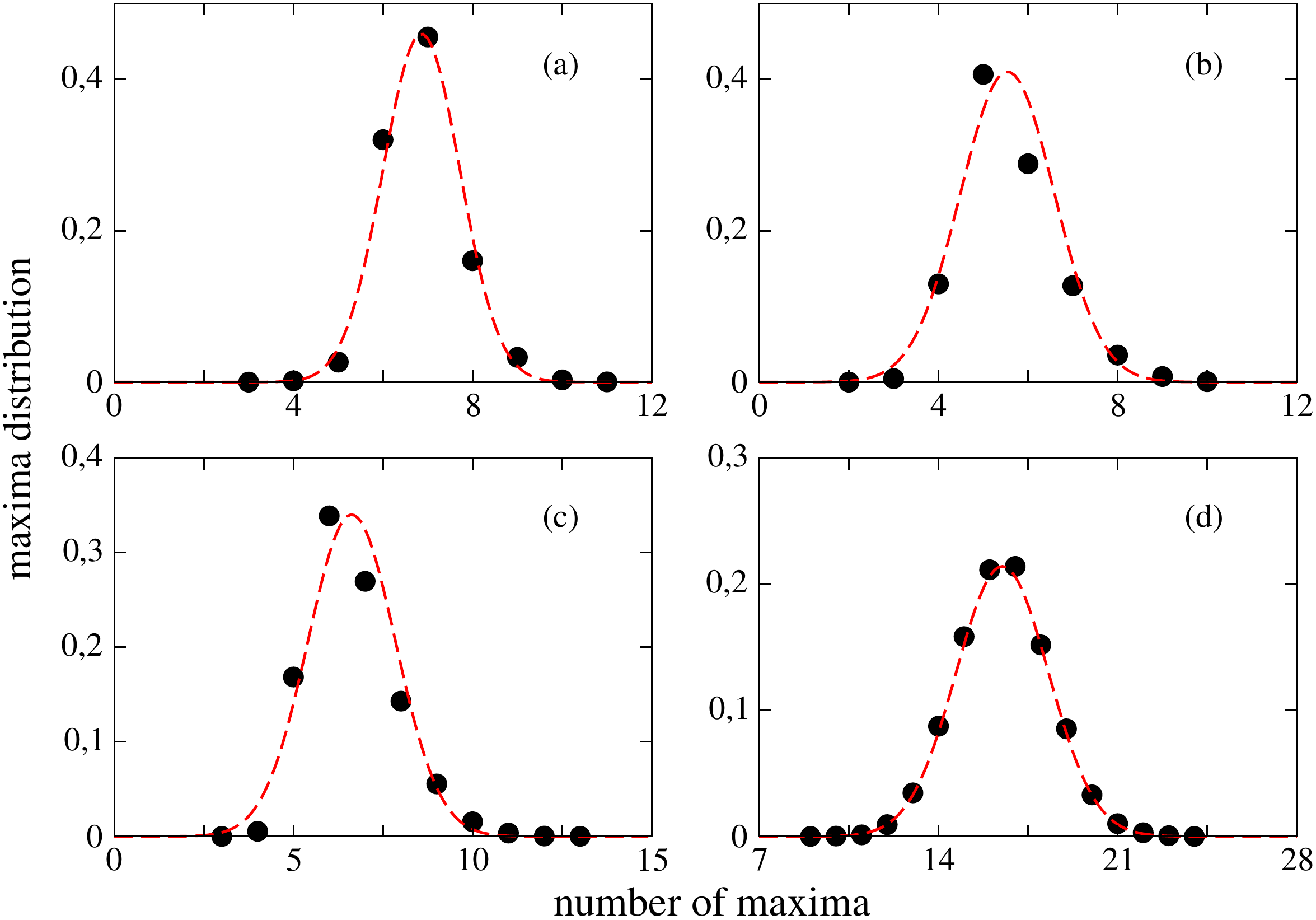}
\caption{The black dots represent the histograms of the number of maxima in the interval of $10^3$ generations and the red dashed lines show the fit as Gaussian curves. The cases (a) and (b) represent the systems $X3$ e $X4$, in which all the species compete cyclically. The figure (c) is for the three species $A$, $B$ e $C$ of the system $SX3$, and the case (d) stands for the the species $D$, the apex predator.}
\label{maximos}
\end{figure}

We provided $3\times 10^3$ realizations for the systems $X3$ and $X4$, and $3\times 10^4$ for the system $SX3$, in order to obtain the average density of maxima $\langle\rho_t\rangle$ in the interval of $10^3$ generations. The Fig. \ref{maximos} shows the histogram normalized for the number of maxima in the interval of $10^3$ generations (black dots) and in Table \ref{tabela_densidade} one shows the values for the average density of maxima and its standard deviations. We use the average and standard deviations values in Table \ref{tabela_densidade} to fit a Gaussian curve, as can be seen in red dashed line in Fig. \ref{maximos}. Note  that the peak distributions have a Gaussian behavior.
\begin{table}[!ht]
\centering
\caption{Average number of maxima in the interval of $10^3$ generations. In the systems $X3$ and $X4$ all the species evolve in cyclic competition, while in the system $SX3$ the species $D$ represents the apex predator.}
\label{tabela_densidade}
\begin{tabularx}{\linewidth}{X|X X X X}
 \hline \hline
 System                   &    $X3$       &    $X4$        &    $SX3$      &    $SX3$    \\ \hline
 Species                 &      $A,B,C  $ &   $A,B,C,D  $  &  $  A,B,C  $  &    $  D $    \\ 
$\langle\rho_t\rangle$   &    $6.858$    &    $5.544$     &    $6.625$   &    $16.496$    \\
$\;\sigma_t$               &    $0.862$    &    $1.052$     &    $1.222$    &    $1.807$    \\
 \hline\hline
 \end{tabularx}
 \end{table}

In order to get further information concerning the dynamical evolution of the systems, we computed the autocorrelation function, which is defined by
\begin{equation}
 C_{xx}(t,t') \equiv \frac{1}{\sigma^2}\langle x(t)x(t')\rangle - \frac{1}{\sigma^2}\langle x(t)\rangle\langle x(t')\rangle
\end{equation}
where $x \in {A,B,C,D}$, $\sigma^2 = \langle x(t)^2 \rangle - \langle x(t) \rangle^2$, and $\langle \cdots \rangle$ is to be understood as an average in the ensamble. In Ref.~\cite{Ramos_2011}, the authors have shown (in a different context) that the density of maxima of an observable that fluctuates can be obtained from its correlation function. This result was recently used \cite{Bazeia_2017} to provide a way to connect the density of maxima with the correlation length of the density of species. The result is obtained with the use of the maximum entropy principle and can be written in the form \cite{Bazeia_2017,Ramos_2011}
\begin{equation}\label{eq:dens}
\langle\rho_t\rangle = \frac{1}{2\pi}\sqrt{-\frac{T_4}{T_2}};\hspace{1cm}  
T_j \equiv \left.\frac{d^j}{d(\delta t)^j}C(\delta t)  \right|_{\delta t=0}.
\end{equation}

\begin{figure}[!ht]
\centering
\includegraphics[width=\linewidth]{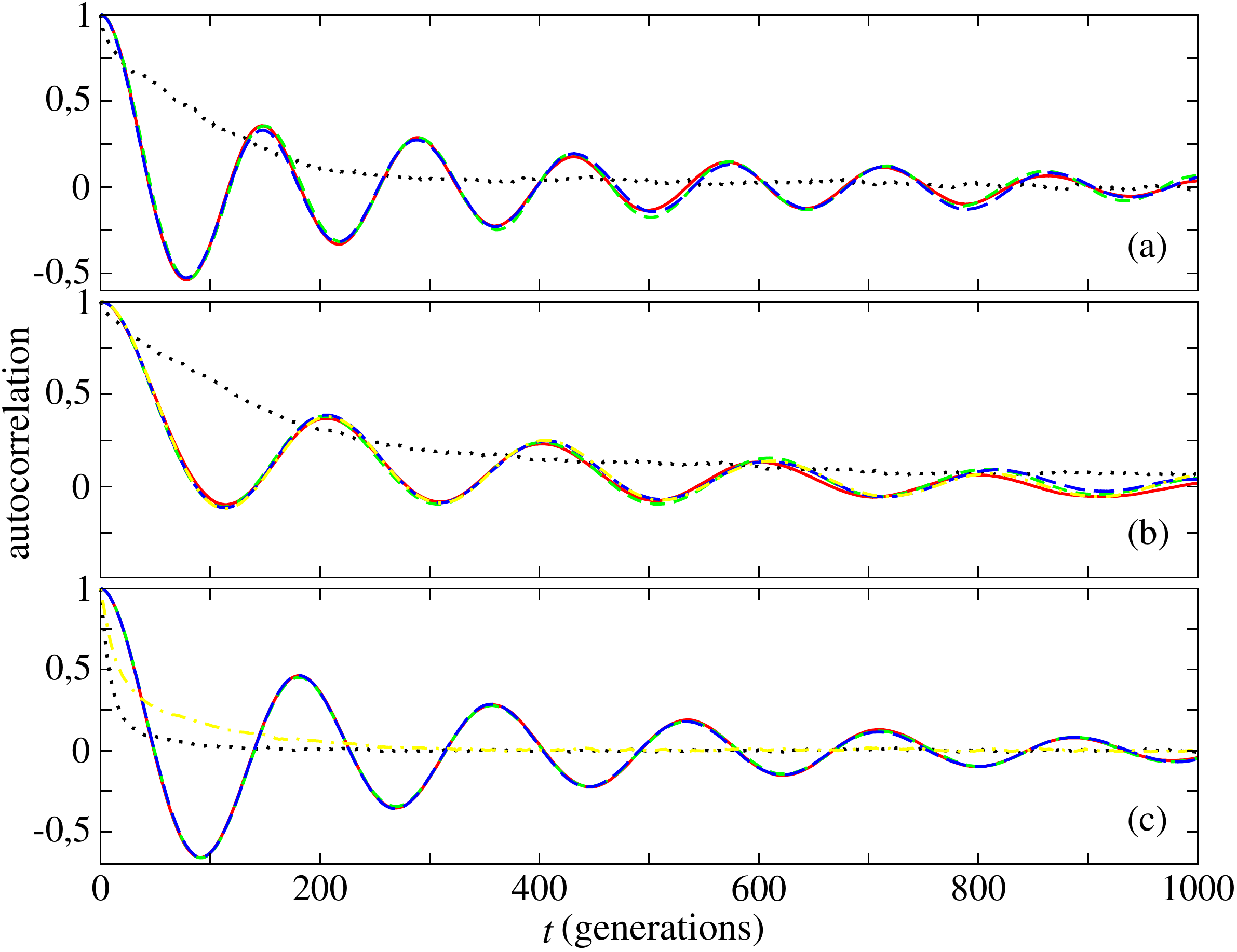} 
\caption{The autocorrelation functions for the systems $X3$ (a), $X4$ (b) and $SX3$ (c). The colors identify the species, and the black dots curve stand for the empty sites. In (c) the three species compete cyclically in the presence of an apex predator, in yellow.}
\label{correlacoes1}
\end{figure}

We then develop stochastic simulations in the square lattice for the three systems $X3$, $X4$ and $SX3$ and show in Fig.~\ref{correlacoes1} the autocorrelation function for each system, there represented by the curves shown in the panels (a), (b) and (c), respectively. The black-dot curves stand for the empty sites, and the colored curves represent the species with their respective colors. One notices that there is strong accord among the results for the species that evolve in cyclic competition, with an oscillating behavior similar to the one found in Ref.~\cite{Bazeia_2017}.

Let us now focus on the species that compete cyclically. One notices from Fig.~\ref{correlacoes1} that for $t$ near the origin, it is possible to map the data of the autocorrelation with
\begin{equation}\label{eq:aprox1}
C(t) = \cos (\omega t + \phi), 
\end{equation}
where $\omega$ and $\phi$ are real parameters. Thus, one uses Eq. (\ref{eq:dens}) to write the average density of maxima for species that evolve under cyclic competition in the form
\begin{equation}
 \langle\rho_t\rangle = \frac{\omega}{2\pi}. 
\end{equation}
However, for the apex predator the initial form of its autocorrelation function can be described by
\begin{equation}\label{eq:aprox2}
C(t) = \exp(a_0 t)\sum^n_{n=1} a_n t^n,
\end{equation}
where $a_0,a_1,a_2,...,$ are real parameters. Since the data are strongly correlated for $n\geq3$, we used $n = 4$ to get the density of maxima for the apex predator. It gives 
\begin{eqnarray}
 \langle\rho_t\rangle = \frac{1}{2\pi}{\sqrt {-{\frac {{{{\it a}_0}}^{4}\!+\!4{{{\it a}_0}}^{
3}{{\it a}_1}\!+\!12{{{\it a}_0}}^{2}{{\it a}_2}\!+\!24{{\it a}_0}{{\it a}_3}
\!+\!24{{\it a}_4}}{{{{\it a}_0}}^{2}+2{{\it a}_0}{{\it a}_1}+2{{\it a}_2}}}}}.\nonumber \\
\end{eqnarray}

\begin{figure}[!ht]
\centering
\includegraphics[width=\linewidth]{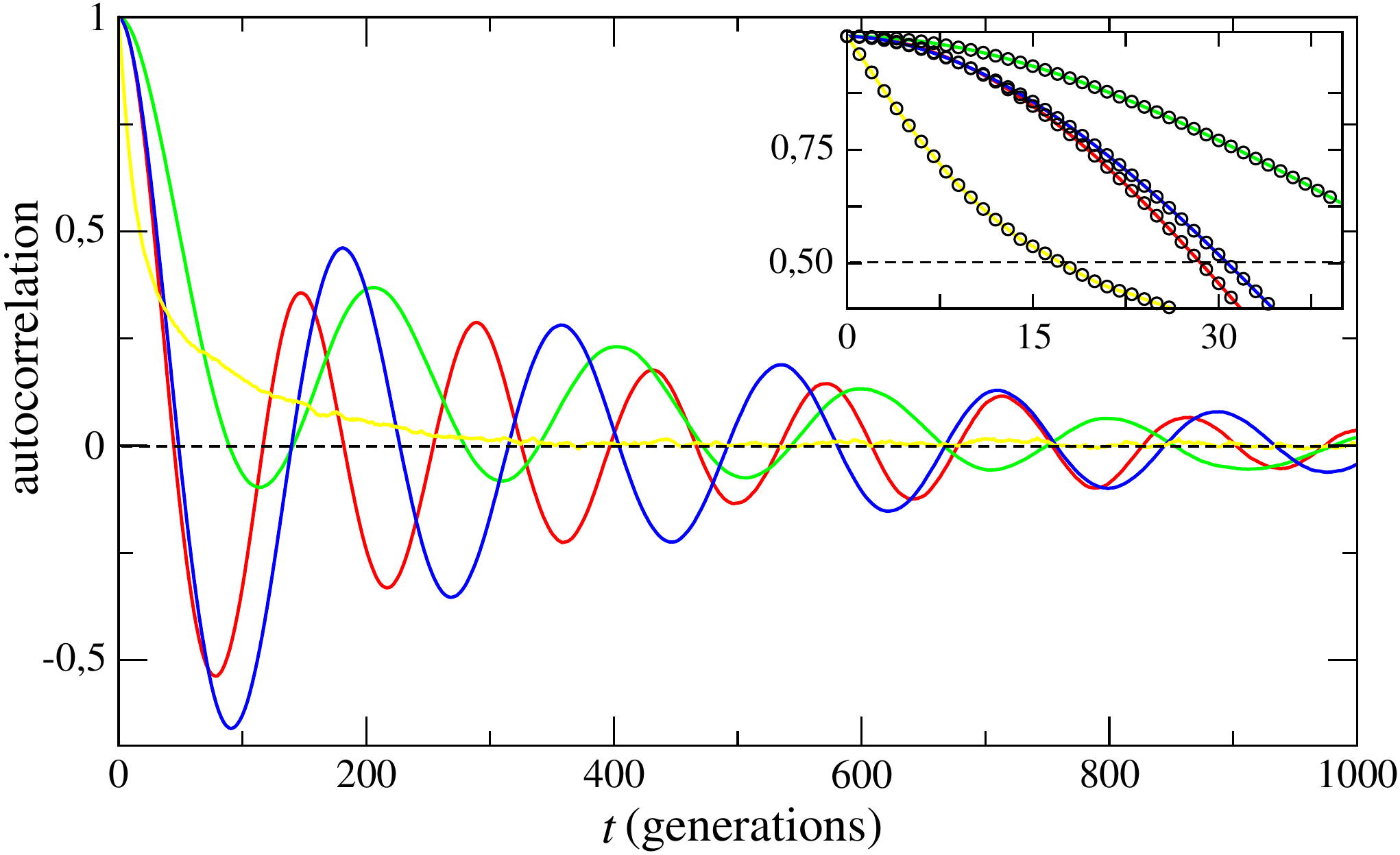} 
\caption{The autocorrelation function for the systems $X3$ (red curve), $X4$ (green curve) and $SX3$ (blue and yellow curves). The yellow curve represents the apex predator. The inset shows perfect agreement between the numerical simulation and the approximations used to fit the curves, shown as the empty-ball curves.}
\label{correlacoes4}
\end{figure}

In Fig.~\ref{correlacoes4} one depicts the autocorrelation functions for the three distinct systems, $X3$, $X4$ and $SX3$, with the red curve representing the $X3$ system, the green curve the $X4$ system, and the blue and yellow curves for one of the three competing species and for the apex predator of the $SX3$ system, respectively. Also, in the inset one displays the same curves together with the corresponding fitting functions as the empty-ball curves that show excellent accord with the colored curves. One recalls that the colored curves result from the stochastic simulations, and the empty-ball curves come from the approximations \eqref{eq:aprox1} and \eqref{eq:aprox2} used above to fit the colored curves. The numerical values are shown in Table \ref{tabela_densidade2}.

\begin{table}[!ht]
\centering
\caption{Average number of maxima in a interval of $10^3$ generations, obtained from the relation \eqref{eq:dens}.}
\label{tabela_densidade2}
\begin{tabularx}{\linewidth}{X|X X X X}
\hline \hline
System                   &    $X3$    &    $X4$       &    $SX3$   &    $SX3$    \\
Species                  & $A, B, C$  & $A, B, C, D$  & $A, B, C$  &    $D$    \\ 
 $\langle\rho_t\rangle$   & $6.081$   &    $4.402$    &    $5.802$ &    $16.537$    \\
 \hline\hline
 \end{tabularx}
 \end{table}

In the systems $X3$, $X4$, and $SX3$ that we have studied, one noticed that species in cyclic competition appear to have the tendency to clusters or agglomerate, as a mechanism to survive in the competing environment. However, the apex predator behaves differently, tending to spread uniformly in the lattice, as it is shown in the right panel of Fig.~\ref{foodwebs} (d). In order to quantify this behavior, one should investigate the number and size of the clusters of the species. To implement this possibility, we used the Hoshen-Kopelman algorithm \cite{HK_1976} to compute the average number of clusters $n_s(t)$ of size $s$ as a function of time. This allows that we introduce the quantities 
\begin{eqnarray}
 N(t) &=& \sum_s n_s(t) \\
 S(t) &=& \frac{\sum_s s^2\, n_s(t)}{\sum_s s\, n_s(t)}
\end{eqnarray}
which represent the number of clusters and the average size of the clusters, respectively. 

\begin{figure}[!ht]
\centering
\includegraphics[width=\linewidth]{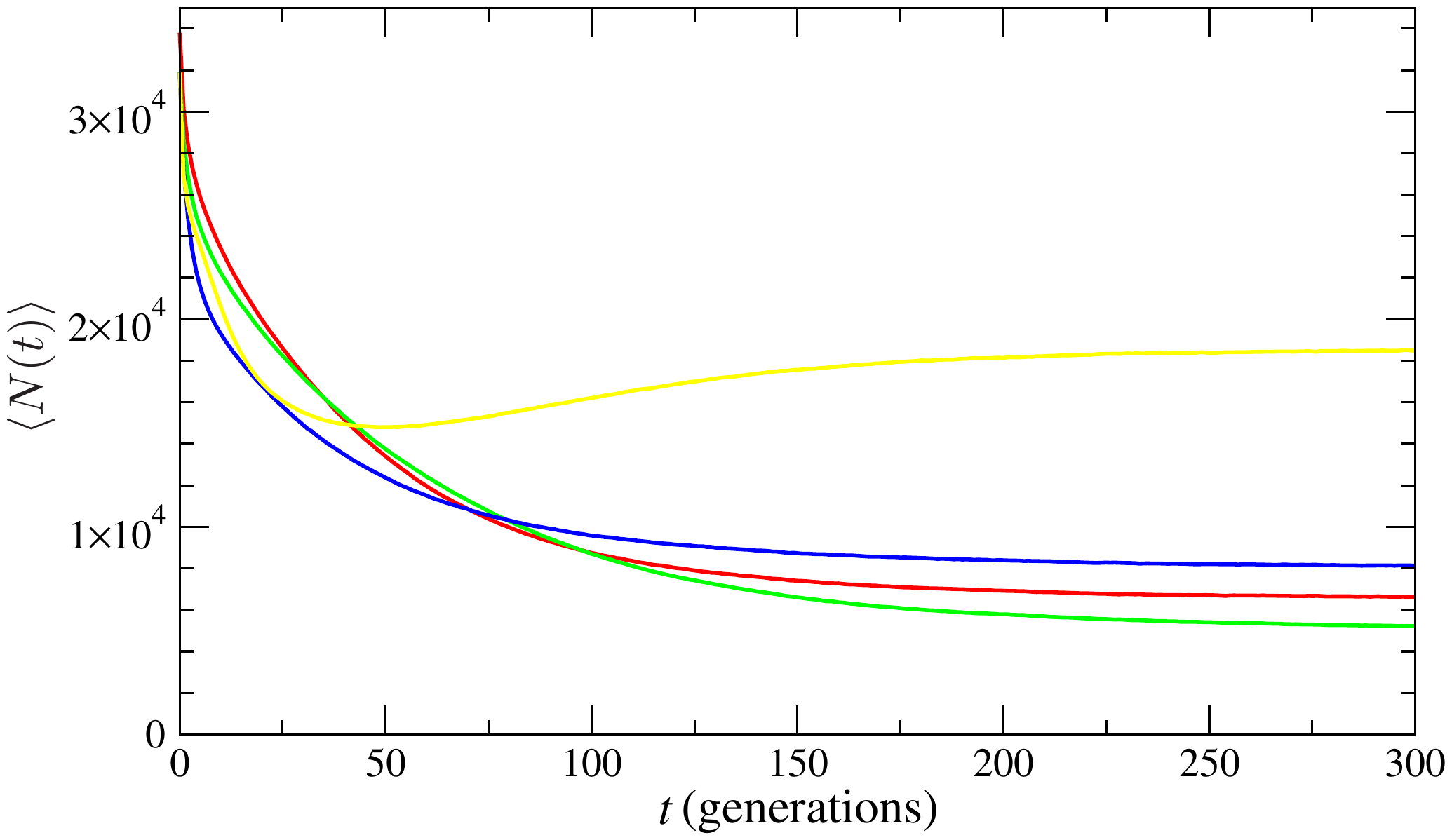}
\caption{Average number of clusters $\langle N(t)\rangle$ as a function of time. The red, green and blue lines represent the species in cyclic competition in the systems $X3$, $X4$ and $SX3$, respectively. The yellow curve stands for the apex predator.} 
\label{media-num-clusters}
\end{figure}

\begin{figure}[!ht]
\centering
\includegraphics[width=\linewidth]{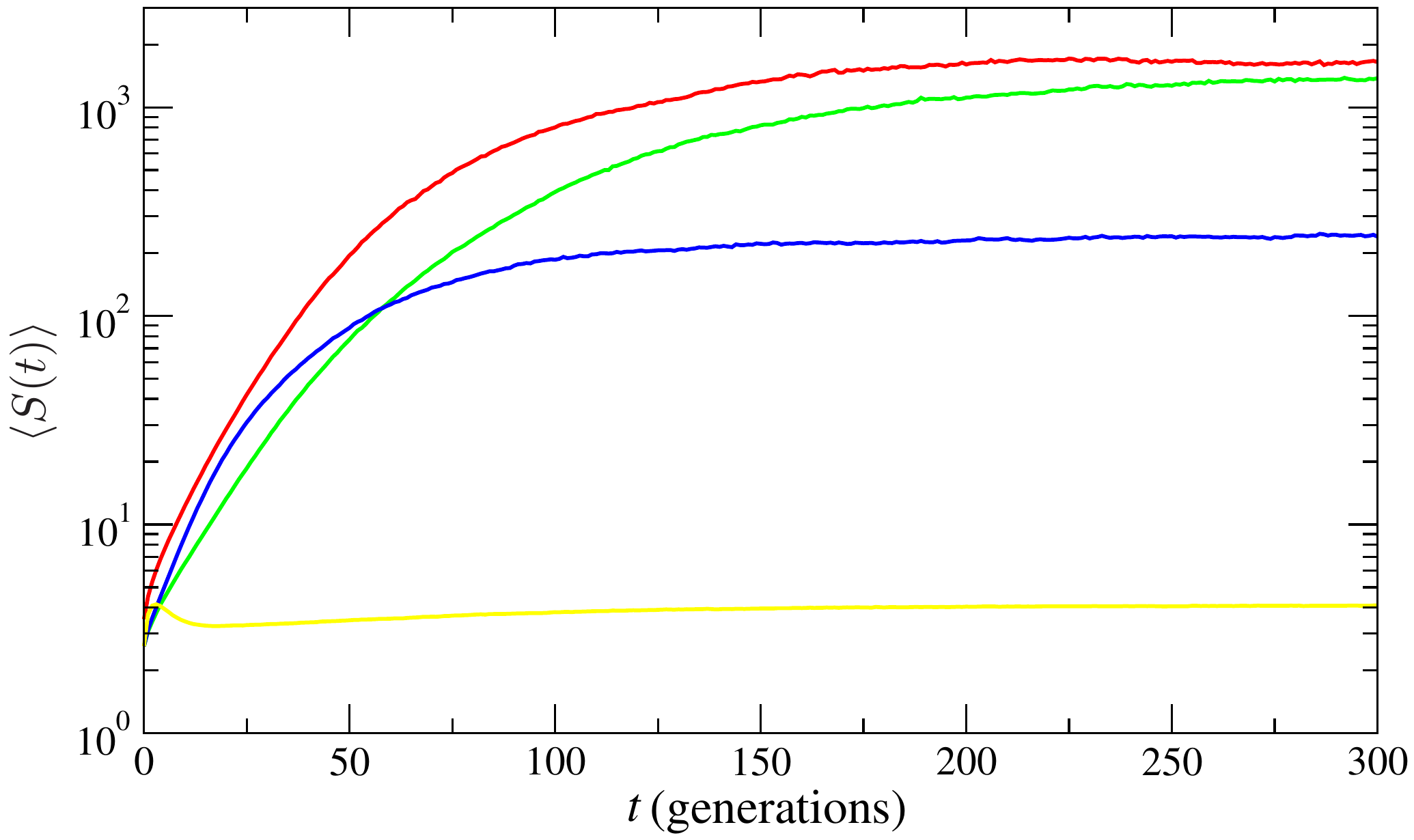}
\caption{The mono-log behavior of the average size of the clusters $\langle S(t)\rangle$ as a function of time. The colors follow as in Fig.~\ref{media-num-clusters}.}
\label{media-tam-clusters}
\end{figure}

We display the results in Figs.~\ref{media-num-clusters} and \ref{media-tam-clusters}, which show the average of the number of clusters and the average size of such clusters as a function of time in the case of 128 realizations, respectively. The results show that the presence of the apex predator contributes to diminish the size of the clusters of the species that compete cyclically, increasing its number. Also, the tendency of the apex predator to spread uniformly in the lattice is kept; as we see from Fig.~\ref{media-tam-clusters}, the average size of the group of apex predators is very small, being $3$ individuals in the simulations shown in the figure.   


\section{Comments and Conclusions}
\label{end}

In this work we studied the presence of an apex predator in a system composed of three distinct species that compete cyclically following the standard rules of reproduction, migration, and predation, which predation controlled as in the paper-rock-scissors game. The apex predator is a superpredator, since it predates all the other species and is not predated by any of them. The population of apex predator does not increase indefinitely because it dies with the ratio controlled by $\beta$, as suggested by the rule \eqref{eq:morte}. 

We studied the system $X3$, which is composed of three distinct species that evolve in cyclic competition, and then two systems of four species, one, the system $X4$, where all the species compete among themselves and have similar behavior, and the other, $SX3$, where three species compete cyclically and the other one represents the apex predator. We followed ideas developed before in the works \cite{Bazeia_2017,HK_1976} to show that when the species evolve under cyclic competition, the abundance or density of species oscillates around an average value, producing maxima and minima. Also, we computed the average number of maxima for the three models and related it with the correlation length of the abundance of species. We also noticed that the presence of the apex predator decreases the amplitude of oscillation of the three species.  

Other results indicated that the autocorrelation function for the abundance of species that evolve under cyclic competition presents a senoidal behavior, but for the apex predator the behavior is exponential. It was also noticed that while the species that compete cyclically tend to cluster and form spiral patterns to survive, the presence of the apex predator diminishes the average size of these clusters, increasing its number in a way that contributes to destroy the spiral patterns without jeopardizing biodiversity. We also found that the apex predator does not form large clusters, preferring to spread out uniformly into the lattice. 

The model for the apex predator studied in this work does not account for several aspects such as the interference among individuals of the same species in the search for the corresponding prey, the conversion of successful predator-prey hunts into new predators, the age of the predators, which may interfere in the predation ratio, and so on. The addition of new rules with focus on specific scenarios is a challenging task that may appear as natural extensions of the current work. Despite the simplicity of the systems studied, the results of this work are of current interest and suggest the need of new, more detailed investigations on the behavior of the apex predator and how it may change basic aspects of the environment that favors biodiversity. Other issues concern investigations related to ordering of heterogeneous systems and complex networks, with the predator-prey Lotka-Volterra mean field theory in the presence of an apex predator and also, with Monte Carlo simulations of n-state Potts models and other complex fluid models in Statistical Physics \cite{landau}. We hope to further report on this in the near future.

\section*{Acknowledgments}
This work is partially supported by The Brazilian agencies CAPES and CNPq.

\end{document}